# Magnetic non-collinear neutron wave resonator

Yu.N. Khaydukov and Yu.V. Nikitenko[*]

*Frank Laboratory of Neutron Physics, JINR, 141980, Dubna, Moscow Region, Russia*

**Abstract**
Equations for the neutron reflection amplitude from a magnetic non-collinear wave resonator (NWR) are obtained. It is shown that resonances of the same reflection order of neutrons experiencing spin-flip (spin-flip neutrons) appear in pairs. Conditions under which in the resonator square enhancement of the spin-flip neutron reflection intensity with respect to growth of the portion of scattered and absorbed neutrons takes place, are determined.

[*]Corresponding author
*E-mail address:* nikiten@nf.jinr.ru (Yu.V. Nikitenko)

## 1. Introduction

A layered structure operating as a neutron resonator (neutron wave resonator (NWR)) is a three-layer structure where the edge layers (the first and the third layer) have a higher neutron-matter interaction potential than the middle (second) one. In the resonator neutron density grows due to multiple neutron reflection and as a result, neutron absorption increases [1]. Therefore, NWR is used to increase the measuring sensitivity of the probability of processes causing neutron absorption [2]. In a magnetic non-collinear NWR where the magnetic induction vectors of layers are non-collinear or the induction vector of some layer is non-collinear with the external magnetic strength vector, an increase in the neutron spin-flip probability in the resonance is stronger than growth of neutron absorption probability [3, 4]. In the present work the neutron spin-flip process in NWR is investigated in detail.

## 2. Non-magnetic NWR

Let neutrons come to NWR from vacuum on the side of the first layer (Fig. 1). To obtain simpler equations, we investigate a NWR whose middle layer is a vacuum space. By solving the Schrodinger equation the neutron wave functions corresponding to the direction forward to the structure (denoted by the index "f") and backward ("b") are presented as [5]

$$\psi_f(y) = \exp(ik(L_2-y))(1- r_1 r_3 \exp(2ikL_2))^{-1} t_1 \psi_0, \quad \psi_b(z) = \exp(2ik(L_2-y))\psi_f(y) \quad (1)$$

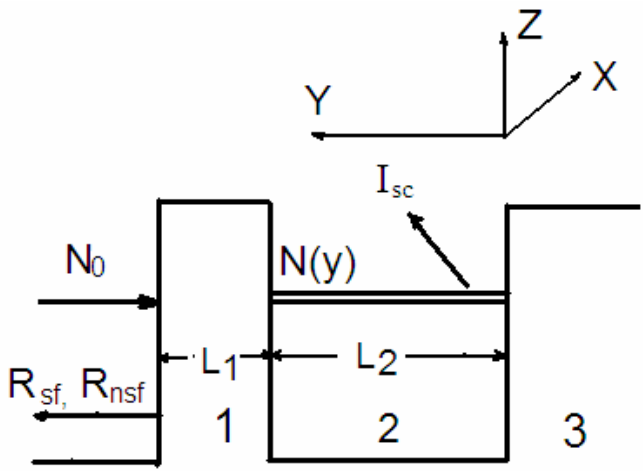

**Fig. 1.** The schematic view of a three-layer NWR.

where $\psi_0$ is the wave function of incident neutrons, $t_1$ is the transmission amplitude for the first layer, $r_1$, $r_3$ are the reflection amplitudes for the first and third layers, respectively, $L_2$ is the thickness of the second layer, $k$ is the neutron wave vector component perpendicular to the layers in the second layer (in vacuum), $y$ is the coordinate in the depth of the structure measured in the backward direction from the interface of the second with the third layer.

From Eq. (1) it follows that the neutron density in the second layer is

$$N = |\psi \equiv \psi_f + \psi_b|^2 = |1+\exp(2ik(L_2-y))|^2 |\psi_d|^2 =$$

$$4\cos^2(k_R(L_2-y))\exp(-k_I(L_2-y))|(1-r_1r_3\exp(2ikL_2))^{-1}|^2|\psi_0|^2, \tag{2}$$

where $k_R$, $k_I$ are the real and imaginary components of the wave vector, respectively.

For the neutron density enhancement coefficient we have

$$E_N = N(r_1,t_1)/N(r_1=0,t_1=1) = |t_1|^2|(1-r_1r_3\exp(2ikL_2))^{-1}|^2 \tag{3}$$

In the resonance, if $\varphi_1 + \varphi_3 + 2k_R L_2 = 2\pi n$, where, $\varphi_1$ and $\varphi_3$ are the phases of the complex reflection amplitudes $r_1$ and $r_3$, respectively, in the absence of neutron absorption in the first layer ($|t_1|^2 = 1-|r_1|^2$), we obtain

$$E_N = (1-|r_1|^2)/(1-|r_1r_3|\exp(-2k_I L_2))^2, \tag{4}$$

In the case of total neutron reflection from the third layer ($|r_3|= 1$), little neutron absorption ($2k_I L_2 \ll 1$) and the absolute amplitude of the first layer $|r_1| \approx 1$, Eq. (4) transforms into a simpler one, i.e.,

$$E_N = (1+|r_1|)/(1-|r_1|\exp(-2k_I L_2)) \tag{5}$$

It is seen that the enhancement coefficient $\eta_N > 1$ and theoretically, it can be infinitely large if $|r_1| \to 1$ and neutron absorption in the second layer is zero ($k_I = 0$). For the absorbed (or scattered) in second layer neutrons we have

$$I_{sc} = 4\int vW\tilde{N}\cos^2(k_R(L_2-y))\exp(-k_I(L_2-y))|\psi_0|^2 dy \tag{6}$$

where $W = E_N\sigma$ is the neutron absorption probability in a layer whose thickness equals unity, $\sigma$ is the neutron absorption cross-section, $\tilde{N}$ is the density of neutron absorption centers in second layer, $v$ is the neutron velocity.

It is seen that the growth of neutron density ($E_N > 1$) results in an increase of the measuring sensitivity $\partial J_{sc}/\partial \sigma$ of the neutron absorption probability. Now, we will show (making use of the Schrodinger equation) that the neutron absorption probability in the resonator can be increased. To this end, we express the neutron absorption cross section in the second layer through the imaginary component of the interaction potential $U_{2I}$ [6]

$$U_{2I} = (h/4\pi)\tilde{N}\sigma v, \tag{7}$$

where h is the Plank constant.

As $U_I$ is small compared to the potentials of the edge layers and the kinetic energy of the neutron, we assume that the amplitudes of neutron reflection from the first layer in both directions are equal ($r_{1f} = r_{1b} \equiv r_1$). The reflection amplitude from the resonator is then

$$r = r_1+t_1^2 r_3\exp(2ikL_2)(1-r_1 r_3\exp(2ikL_2))^{-1}, \tag{8}$$

where $k = k_R+ik_I =(k^2+iU_{2I})^{1/2} = (k^4+U_{2I}^2)^{1/4}\exp(i\chi/2)$, $tg(\chi) =U_{2I}/k^2$, $U_{2R}=0$.
In the resonance, for $\varphi_1+ \varphi_3 +2k_R L_2 = 2\pi n$ we have

$$r = [r_1+r_3(t_1^2-r_1^2)\exp(2ikL_2)]/[1-|r_1 r_3|\exp(-2ik_I L_2)] \tag{9}$$

Making use of the condition $(t_1^2-r_1^2) = -\exp(2\varphi_1)$ [7] and again of the condition $\varphi_1+ \varphi_3 +2k_R L_2 = 2\pi n$ we obtain

$$r = |r_1|\exp(2i\varphi_1)[|r_1|-|r_3|\exp(-2k_I L_2)]/[1-|r_1 r_3|\exp(-2k_I L_2)]/r_1 \tag{10}$$

In the case of total neutron reflection from the third layer of the resonator ($|r_3|=1$) the neutron absorption coefficient $M =1-|r|^2$ is

$$M = 4k_I L_2(1+|r_1|)/(1-|r_1|\exp(-2k_I L_2)) \tag{11}$$

From Eq. (11) it follows that the neutron absorption enhancement coefficient

$$E_M=M(r_1)/M(r_1=0)=(1+|r_1|)/[1-|r_1|\exp(-2k_I L_2)] \tag{12}$$

Comparing Eq. (12) with Eq. (4) we obtain

$$E_N=E_M \tag{13}$$

Let us exactly calculate some resonator structures made of real materials. There follow the values of the real $U_R$ and the imaginary $U_I$ components of the nuclear potentials of the materials we use, namely, copper $U_R(Cu)=1$, $U_I(Cu)=-1.45E-04$, beryllium $U_R(Be)=1.44$, $U_I(Be)=-4.3E-06$, aluminum $U_R(Al)=0.32$, $U_I(Al)=-6.4E-06$, iron $U_R(Fe)=1.23$ and $U_I(Fe)=-5E-05$, and cobalt $U_R(Co)=0.625$ and $U_I(Co)=-2.3E-03$. To pass on to absolute values of the potentials, one should have in mind that in this case the unity ($U_R(Cu)$) is 172 neV. Figure 2 shows the results of calculation of the functions $N(k)$ and $M(k)$ for the structure $Be(L_1)/Al(L_2=80nm)/Be$ for $L_1 = 0$, 10, 20 and 30 nm. It is seen that with increasing $L_1$ (proportionally increasing $|r_1|$) the neutron density and the neutron absorption coefficient grow at the resonance values of the wave vector $k(n=1)=0.00603$Å$^{-1}$ and $k(n=2)=0.00802$Å$^{-1}$. At the same time, the enhancement coefficient $E_N(L_1=30nm)$ is 254.9 and 87.8 in the resonances of the first ($n=1$) and second ($n=2$) order, respectively. In its turn, the absorption enhancement coefficient $E_M(L_1=30nm)$ is 255.3 and 88.0 in the first and the second resonances, respectively. It is seen that the equality $E_N=E_M$ holds quite well. A little larger $E_M$ than $E_N$ is due to neutron absorption in the first beryllium layer for $L_1=30$ nm. It should be remembered that we obtained $E_N=E_M$ under assumption that neutrons are not absorbed in the first and the second layer, i.e., the imaginary potential of these layers was set to be zero.

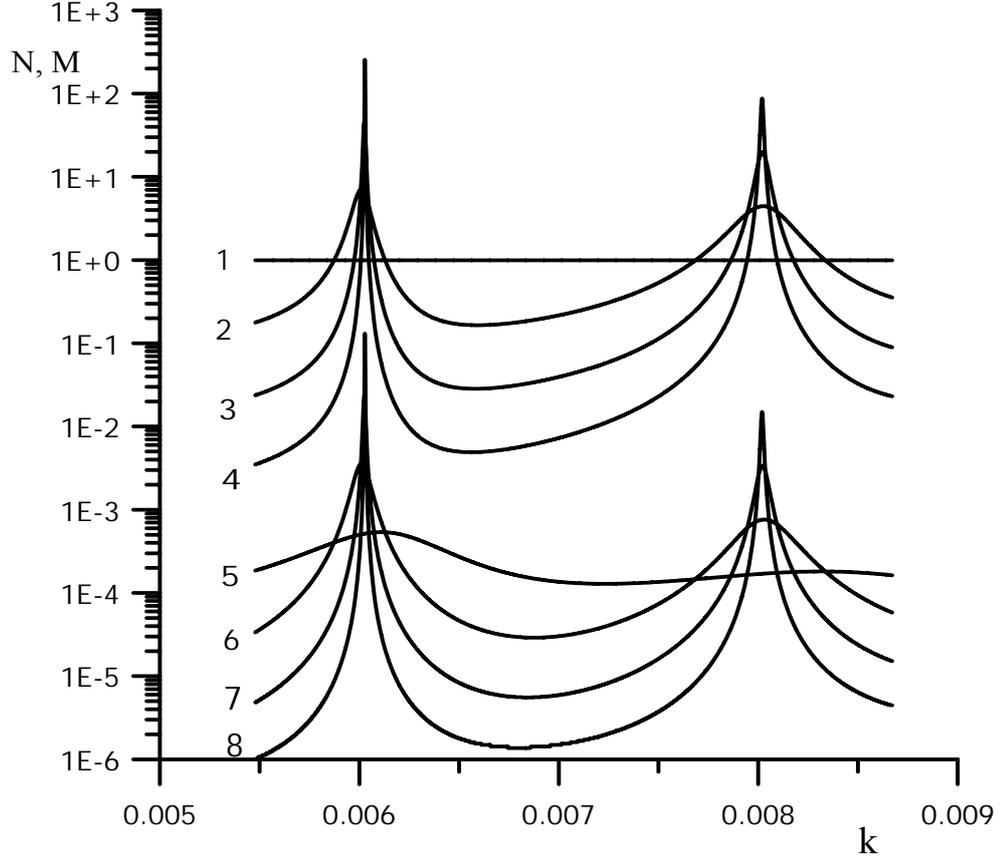

**Fig. 2.** The neutron density $N$ (curves 1-4) and neutron absorption coefficient $M$ (curves 5-8) vs the perpendicular component of the wave vector $k(\text{Å}^{-1})$ for the structure $Be(L_1)/Al(L_2=80nm)/Be$ at different $L_1$: 0 – curves 1,5; 10nm – curves 2,6; 20nm - curves 3,7; 30 nm - curves 4,8.

### 3. Magnetic non-collinear NWR

Let us consider the simplest, though reflecting all the peculiarities, case of a magnetic NWR in which the first layer is magnetic, the second with the length $L_2$ is vacuum, the third is a semi-infinite medium whose magnetic induction vector $\boldsymbol{B}$ is non-collinear with the neutron polarization vector $\boldsymbol{P}$ directed along the axis Z. The external magnetic field is assumed to be zero. Since the spin part of the neutron wave function is a spinor, the amplitude of neutron reflection from the third magnetic layer, $\acute{r}_3$, and consequently, that of neutron reflection from the NWR, $\acute{r}$, are operators [7, 8]

$$\acute{r}_3(\boldsymbol{\sigma B}) = [k+(k^2-U_3-2\boldsymbol{\sigma B})^{1/2}]^{-1}[k-(k^2-U_3-2\boldsymbol{\sigma B})^{1/2}], \tag{14}$$

where $U_3$, $\boldsymbol{B}$ is the interaction potential and the magnetic induction vector in the third layer, respectively, $\boldsymbol{\sigma}$ is the vector of the Pauli matrices $\sigma_x, \sigma_y,$ and $\sigma_z$.

Let us write $\acute{r}_3(\boldsymbol{\sigma B})$ in the Pauli matrix-linear form [7]

$$\acute{r}_3 = r_3^+ + \boldsymbol{b\sigma} r_3^-, \tag{15}$$

where $r_3^+ = 0.5(r^+ + r^-)$, $r_3^- = 0.5(r^+ - r^-)$, $r^+ = [k-(k^2-U_3-B)^{1/2}]/[k+(k^2-U_3-B)^{1/2}]$, $r^- = [k-(k^2-U_3+2B)^{1/2}]/[k+(k^2-U_3+2B_3)^{1/2}]$, $\boldsymbol{b} = \boldsymbol{B}/B$, $B \equiv |\boldsymbol{B}|$.

We substitute Eq. (15) into Eq. (8) and use the technique for transformation of equations containing Pauli matrices [7]. As a result we obtain.

$$\acute{r}=r_1+t_1^2\exp(2ikL)\{r_3^+ + \exp(2ikL)r_1[(r_3^-)^2-(r_3^+)^2]+r_3^-[1-2\exp(2ikL)r_1r_3^+]\boldsymbol{b\sigma}\}/G, \tag{16}$$

where $G = [1-\exp(2ikL)r_1r^-] [1-\exp(2ikL)r_1r^+]$

Let the magnetic induction vector **B** lie in the plane ZX. Then, for the spin-flip amplitude of reflection we have

$$r_{sf} = <\xi^-| \dot{r} |\xi^+> = <\xi^+| \dot{r} |\xi^-> = b_x\exp(2ikL_2) t_1^2 r_3^-(1-2\exp(2ikL) r_1 r_3^+)/ [(1-\exp(2ikL_2)r_1r^-)(1-\exp(2ikL_2)r_1r^+)] \quad (17)$$

where $|\xi^+> = \begin{pmatrix} 1 \\ 0 \end{pmatrix}$, $|\xi^-> = \begin{pmatrix} 0 \\ 1 \end{pmatrix}$

It is seen that $r_{sf}$ has two maximums at the resonance values of the wave vector $k^-$ and $k^+$ satisfying the relation

$$2k^-L_2+\varphi_1+\varphi^- =2\pi n \quad , \quad\quad\quad 2k^+L_2+\varphi_1+\varphi^+ =2\pi n , \quad (18)$$

where $n =0, 1, 2…$; $\varphi^-$ and $\varphi^+$ are the phases of the complex reflection amplitudes $r^-$ and $r^+$, respectively.

For the enhancement coefficient of the spin-flip reflection coefficient $R_{sf} =k_f|r_{sf}|^2/k_i$, where $k_i$ and $k_f$ are the initial and final neutron wave vectors, respectively, we obtain

$$E_{Rsf} = R_{sf}/R(r_1=0, t_1=1) = |t_1^2(1-2\exp(2ikL_2)r_1r_3^+)/ [(1-\exp(2ikL_2)r_1r^-)(1-\exp(2ikL_2)r_1r^+)]|^2 \quad (19)$$

Let us consider the question of relation between the difference of resonance wave vector values $\Delta k = k^- - k^+$ and the half-width of resonances $\delta k^-$ and $\delta k^+$. For $\delta k^-$ and $\delta k^+$ we have

$$\delta k^- = (1-|r_1r^-|)/(2|r_1r^-|^{1/2}L_2), \quad\quad \delta k^+ = (1-|r_1r^+|)/(2|r_1r^+|^{1/2}L_2) \quad (20)$$

In the case of total reflection of neutrons in two spin states ($|r^-|= |r^+|= 1$) the half-widths are equal. They are

$$\delta k^- = \delta k^+ = \delta k = (1-|r_1|)/(2L_2|r_1|) \quad (21)$$

For the difference $\Delta k$ we obtain, using Eq. (18), the following relation

$$\Delta k = (\varphi^+ - \varphi^-)/2L_2 \quad (22)$$

where $\operatorname{tg}(\varphi^+) = ((U_{3R}+2B)/k^2 -1)^{1/2}$, $\operatorname{tg}(\varphi^-) = ((U_{3R}-2B)/k^2 -1)^{1/2}$, $U_{3R}$ is the real part of the interaction potential of the third layer.

If $(U_{3R}, k^2) >> B$, for the resonance of the first order ($n= 1$) we have

$$\Delta k = (\varphi^+ - \varphi^-)/2L_2 = (B/U_{3R}^{1/2})/(kL_2) = (B/U_{3R}^{1/2})/\pi \quad (23)$$

In the case of overlapping resonances, i.e., $\Delta k \le (\delta k^- + \delta k^+)$ we obtain

$$B \le U_{3R}^{1/2}k(1-|r_1|)/|r_1| \quad (24)$$

which relates the magnetic field induction with the interaction potential.

For $U_{3R} \approx k^2$ we have $B \le U_{3R}(1-|r_1|)/|r_1|$. If the condition (24) holds, the enhancement coefficient of overlapped resonances (index "o")

$$E_{\text{Rsf-o}} \approx (1-|r_1|^2)^2/(1-|r_1|)^4 = (1+|r_1|)^2/(1-|r_1|)^2 \tag{25}$$

A comparison of Eqs. (4), (12), and (24) yields

$$E_{\text{Rsf-o}} = E_M^2 = E_N^2 \tag{26}$$

If $\Delta k \to 0$, the half-width of the resonance formed as a result of overlapping of two resonances is

$$\delta K = (1-|r_1|)^2/8L_2 = L_2 \delta k^2/2 \tag{27}$$

The half-width of the resonance determines the time of life of the neutron in the resonator in accordance with the relation

$$\tau = 1/(v \delta K) \tag{28}$$

where $v$ is the neutron velocity.
Thus, it follows that if resonances overlap, spin-flip neutrons live longer in the resonator than those that do not experience a spin-flip. This is explained by the fact that spin-flip neutrons live in two spin states in turn.
Next, let us investigate the question of enhancement of the density of nonspin-flip and spin-flip neutrons in the second layer. Nonspin-flip neutrons are the neutrons in the initial spin state. Therefore, their density is equal to that of neutrons in a nonmagnetic structure, i.e., $E_{\text{Nnsf}} = E_N$. Similarly, in the case of nonoverlapping resonances (index "no") for spin-flip neutrons we have $E_{\text{Nsf-no}} = E_N$. For overlapping resonances the spin-flip neutron density enhancement coefficient is

$$E_{\text{Nsf-o}} = E_{\text{Rsf-o}}/|t_1|^2 = (1-|r_1|^2)/(1-|r_1|)^4 = (1+|r_1|)/(1-|r_1|)^3 = E_{\text{Nnsf}}^3/(1+|r_1|)^2 = E_N^3/(1+|r_1|)^2 \tag{29}$$

Consequently, in the case of overlapping resonances, the spin-flip neutron density in the second layer of the resonator is proportional to cubic neutron density in a nonmagnetic resonator. Figure 3 presents the results of exact calculation of the functions $N_{nsf}(k)$, $N_{sf}(k)$, $R_{sf}(k)$ and $M(k)$ for the structure $Be(L_1)/Al(L_2 = 80nm)/Fe(J_x=10^{-5}T)$ for $L_1=0$ and 30 nm. For example, for the resonance of the first order the enhancement coefficient $E_{\text{Nnsf}}=247.96$, $E_M=248.28=1.0013 E_{\text{Nnsf}}$, $E_{\text{Rsf}}=61484=1.0011 E_{\text{Nnsf}}^2$, $E_{\text{Nsf}}=448719= E_{\text{Nnsf}}^3/3.4$. It is seen that Eqs. (13), (26), (29) hold well. Figure 4 also shows the functions $N_{nsf}(k)$, $N_{sf}(k)$, $R_{sf}(k)$ $M(k)$ for the structure $Be(L_1)/Al(L_2= 80nm)/Fe(J_z=1T, J_x=10^{-5}T)$ being different by the fact that its third layer has a large longitudinal component of induction $J_z=1T$. It is seen that in the functions $N_{sf}(k)$ and $R_{sf}(k)$ there appear pairs of maximums corresponding to pairs of resonances. At the same time, enhancement decreases compared to the previous case. Figure 5 illustrates the case of the resonator structure in an external magnetic field. It is seen that as the magnetic field increases the resonances split and enhancement of the reflection coefficient goes down. The Table summarizes values of the enhancement coefficient $\eta$ and the effective half-width $\delta k_{\text{eff}} = 2L_2 \delta k$ as well as of the parameter $D=E/\delta k_{\text{eff}}$ and $P=E \delta k_{\text{eff}}$ for the case when we measure all the parameters $N$, $M$, $N_{nsf}$, $N_{sf-no}$, $N_{sf-o}$, $R_{sf-no}$, $R_{sf-o}$. The newly introduced parameter $D$ characterizes the measuring sensitivity and the parameter $P$ characterizes the integral over wave vector value of the measured parameter in the resonance. It is seen that $P$ and $D$ are the largest when the spin-flip density is measured in the case of resonance overlapping $N_{\text{sf-o}}$.

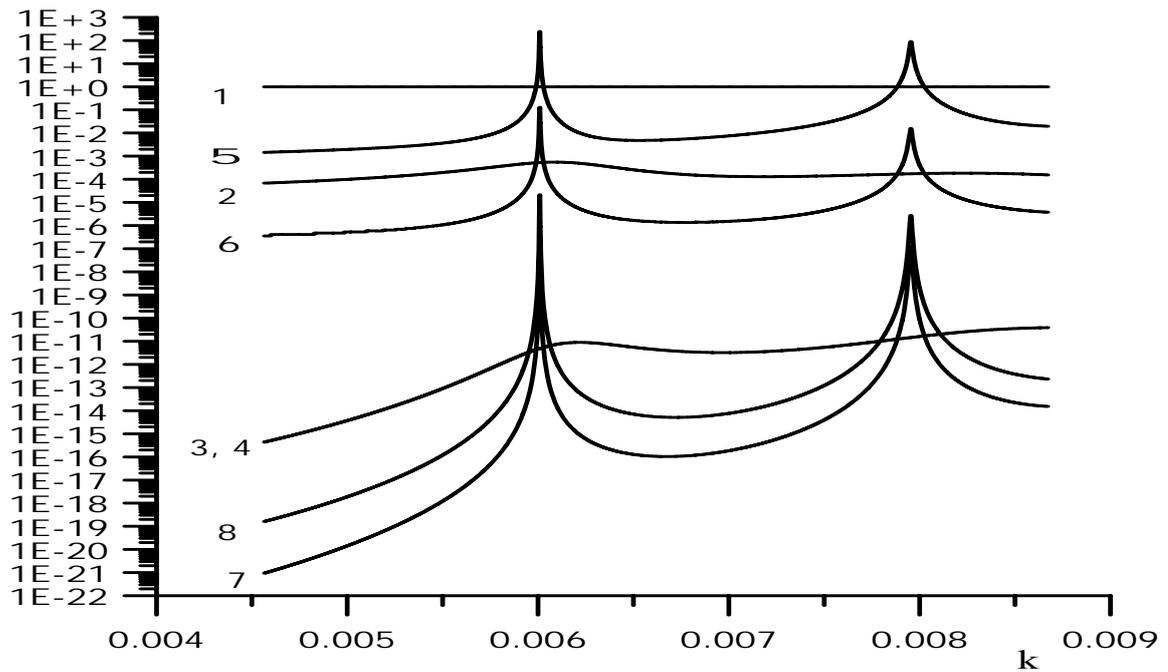

**Fig. 3.** The dependence of $N_{nsf}$(curves 1 and 5), $M$ (curves 2 and 6), $R_{sf}$ (curves 3 and 7) and $N_{sf}$ (curves 4 and 8) on the perpendicular component of the wave vector $k$(Å) for the structure $Be(L_1)/Al(L_2 = 80nm)/Fe(J_x=10^{-5}T)$ at different $L_1$: 0- curves 1, 2, 3, 4 ; 30nm - curves 5, 6, 7, 8.

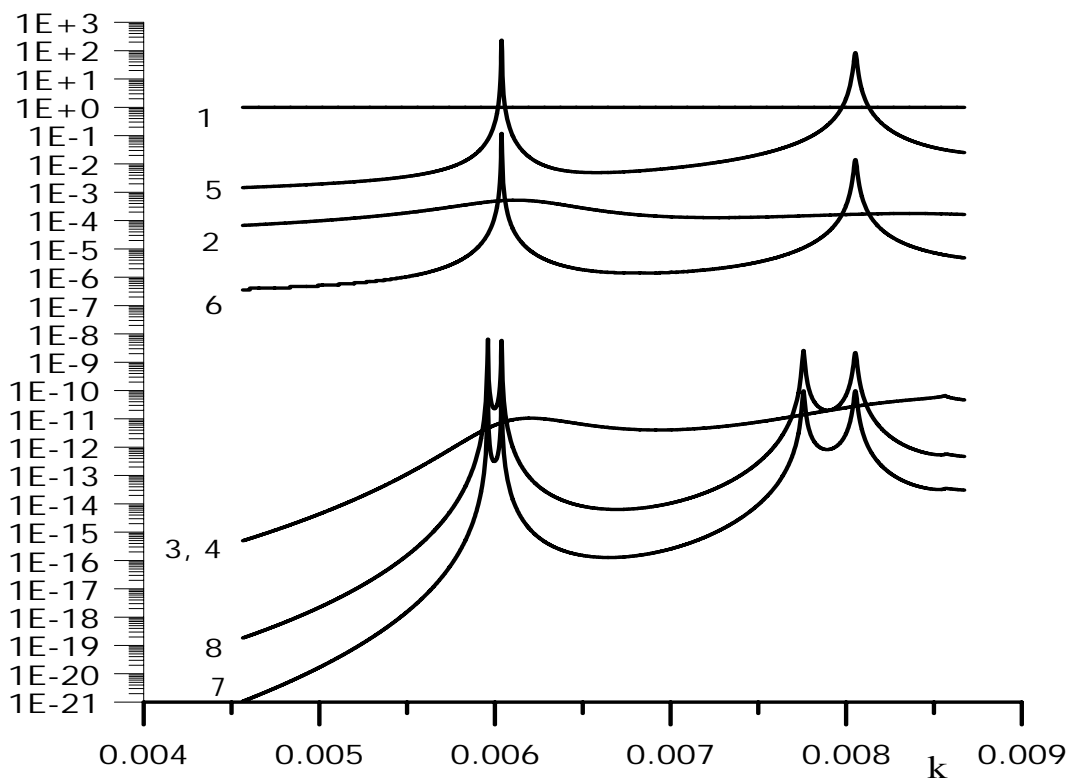

**Fig. 4.** The dependence of $N_{nsf}$(curves 1 and 5), $M$ (curves 2 and 6), $R_{sf}$ (curves 3 and 7) and $N_{sf}$ (curves 4 and 8) on the perpendicular component of the wave vector $k$(Å) for the structure $Fe(L_1)/Al(L_2=80nm)/Fe(J_z=1T, J_x=10^{-5}T)$ at different $L_1$: 0 - curves 1, 2, 3, 4 ; 30nm - curves 5, 6, 7, 8.

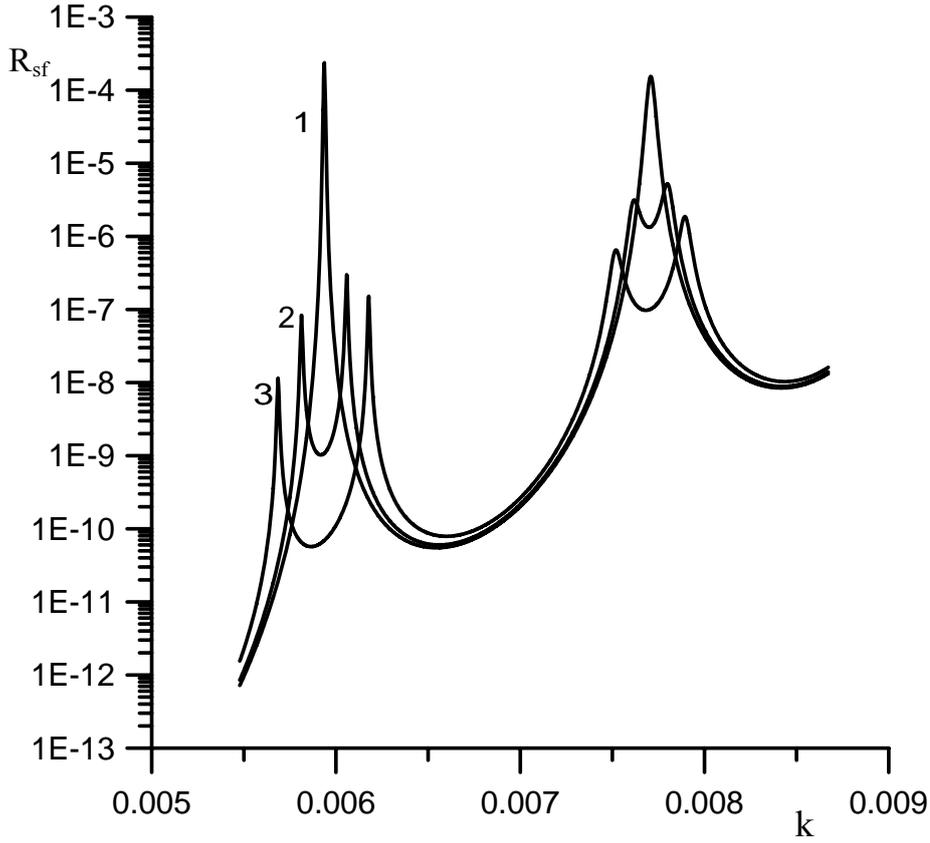

**Fig. 5.** The dependence of $R_{sf}$ on the perpendicular component of the wave vector $k(\text{Å}^{-1})$ for the structure $Cu(30nm)/Al(80nm)/Fe(J_x=10^{-3}T)$ for different $H_z$: $10^{-2}$ kOe – curve 1, 0.5kOe – curve 2, 1 kOe – curve 3.

**Table.** The values of $E$, $\delta k_{eff}$, $D$ and $P$ for the measured parameters $N$, $M$, $N_{nsf}$, $N_{sf}$ and $R_{sf}$ for overlapping and non-overlapping resonances.

| Measurable parameter | $E$ | $\delta k_{eff}$ | $D = E/\delta k_{eff}$ | $P = E \delta k_{eff}$ |
|---|---|---|---|---|
| $N, M, N_{nsf}, N_{sf\text{-}no}$ | $(1+|r_1|)/(1-|r_1|)$ | $(1-|r_1|)$ | $(1+|r_1|)/(1-|r_1|)^2$ | $(1+|r_1|)$ |
| $R_{sf\text{-}no}$ | $(1+|r_1|)^2$ | $(1-|r_1|)$ | $(1+|r_1|)^2/(1-|r_1|)$ | $(1+|r_1|)^2(1-|r_1|)$ |
| $R_{sf\text{-}o}$ | $(1+|r_1|)^2/(1-|r_1|)^2$ | $(1-|r_1|)^2/4$ | $4(1+|r_1|)^2/(1-|r_1|)^4$ | $(1+|r_1|)^2/4$ |
| $N_{sf\text{-}o}$ | $(1+|r_1|)/(1-|r_1|)^3$ | $(1-|r_1|)^2/4$ | $4(1+|r_1|)/(1-|r_1|)^5$ | $(1+|r_1|)/[4(1-|r_1|)]$ |

Next, let us investigate a five-layer structure (Fig. 6) which is a nonmagnetic NWR inside whose second layer there is introduced a magnetic layer so that in the five-layer structure the magnetic layer will be a third in order. It should be noted that the structures of the type, i.e., the ones in which the second and fourth layers are superconducting are, for example, used in studies of the interaction between magnetic and superconducting order parameters [10]. We will not repeat the entire sequence of transformations but will simply write the final equation for the spin-flip reflection amplitude

$$r_{sf} = b_x t_1^2 \exp(2ikL_2)(r_{3-}+F/A)/E \qquad (30)$$

where $F = r_{45}[t_{3-}(1-r_{45}r_{3+})t_{3+}+r_{45}r_{3-}t_{3+}^2+t_{3-}t_{3+}(1-r_{45}r_{3+})+t_{3-}^2 r_{45}r_{3-}]$, $A = (1-r_3(B)r_{45})(1-r_3(-B)r_{45})$, $E = [1-r_1\exp(2ikL_2)(r_{3+}+C/A-(r_{3-}+F/A))][1-r_1\exp(2ikL_2)(r_{3+}+C/A+(r_{3-}+F/A))]$,
$C = r_{45}[t_{3+}^2(1-r_{45}r_{3+})+t_{3-}r_{45}r_{3-}t_{3+}+t_{3-}^2(1-r_{45}r_{3+})+r_{45}r_{3-}t_{3+}t_{3-}]$, $r_{45} = \exp(2ikL_4)r_5$,
$r_{3\pm} = 0.5(r_3(B) \pm r_3(-B))$, $t_{3\pm} = 0.5(t_3(B) \pm t_3(-B))$, $r_3(\pm B) = [r(\pm B)(1-e^2(\pm B)/(1-r^2(\pm B)e^2(\pm B))]$,
$t_3(\pm B) = e(\pm B)[(1-r^2(\pm B))/(1-r^2(\pm B)e^2(\pm B))]$, $e(\pm B) \equiv \exp(ik_3(\pm B)L_3)$,
$r(\pm B) = (k-k(\pm B))/(k+k(\pm B))$, $k(\pm B) = (k^2-U_3-(\pm 2B))^{1/2}$.

Within the limits of a sufficiently thin magnetic layer and $k^2 \gg U_3$ when $(r_{3+}, r_{3-}) \ll (t_{3+}, t_{3-})$ the equation for $r_{sf}$ transforms into

$$r_{sf} = 2b_x t_1^2 t_{3-} t_{3+} r_5 \exp(2ik(L_2+L_4))/\{[1-r_1 r_5 \exp(2ik(L_2+L_4))(t_{3+}+t_{3-})^2] [1-r_1 r_5 \exp(2ik(L_2+L_4))(t_{3+}-t_{3-})^2]\} \quad (31)$$

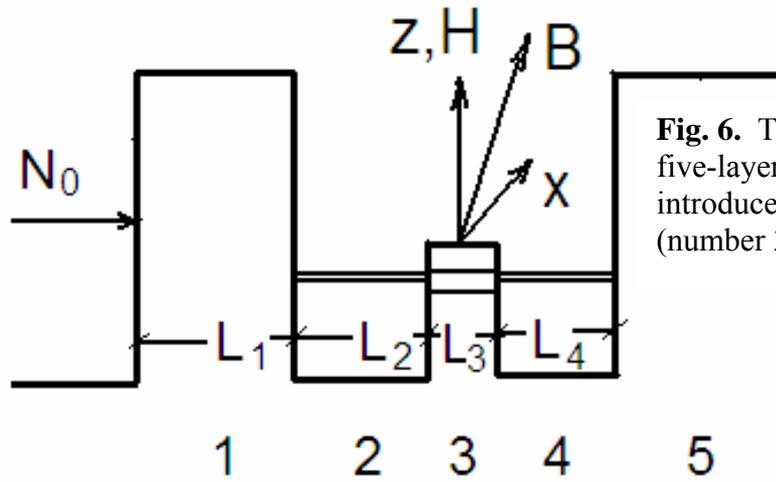

**Fig. 6.** The schematic view of a five-layer NWR with an introduced magnetic layer (number 3).

Looking at the denominator we can say that in this case resonances are realized in pairs. Then for $\Delta k$ we have

$$\Delta k = 2BL_3/[(k^2-U_{3R})^{1/2}(L_2+kL_3/(k^2-U_{3R})^{1/2}+L_4)] \quad (32)$$

From Eq. (32) it follows that $\Delta k$ depends on the relationship between neutron wave phases in the second, third, and the fourth layer of the resonator $kL_3/[(k^2-U_{3R})^{1/2}(L_2+L_4)]$. If $k^2 > U_{3R}$, in the resonance of the first order $(k(L_2+L_3+L_4) \approx \pi)$ we have

$$\Delta k \approx 2BL_3/\pi \quad (33)$$

It is seen that in this case $\Delta k$ is about $(L_2+L_3+L_4)/2L_3$ times smaller than $\Delta k$ that follows from Eq. (23). Figure 7 shows the dependence $R_{sf}(k)$ for the structures $Be(30nm)/Al(19.95nm)/Co(0.1nm, J_z=1T, J_x=10^{-2}T)/Al(59.95nm)/Be$ and $Be(30nm)/Al(19nm)/Co(2nm, J_z=1T, J_x=10^{-2}T)/Al(59nm)/Be$. It can be seen that with increasing thickness of the magnetic layer and increasing $\Delta k$ there appear pairs of resonances. For comparison, there are shown the dependences (curves 3 and 4) corresponding to the layers that are not situated in the resonator.

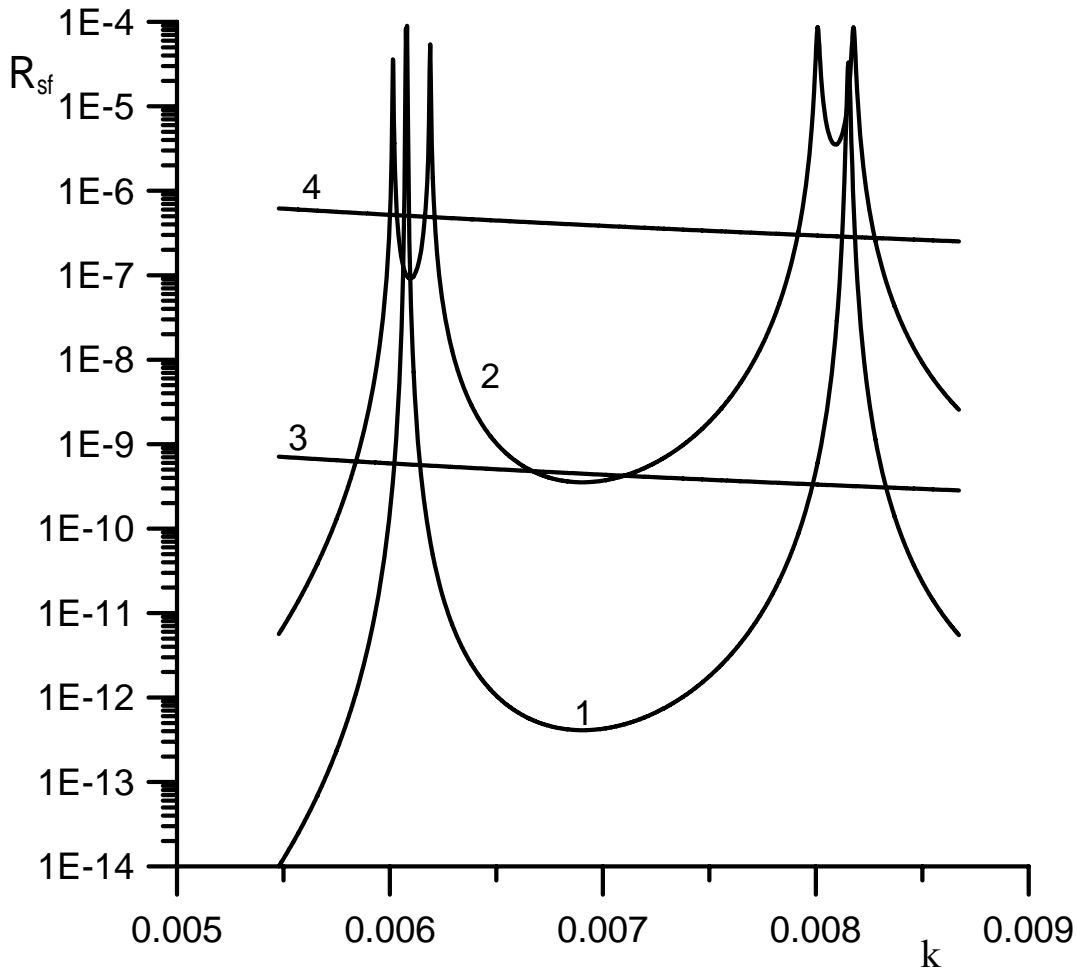

**Fig. 7.** The dependence of $R_{sf}$ on the perpendicular component of the wave vector k(Å$^{-1}$): curve 1 – the structure *Be*(30nm)/*Al*(57nm)/*Co*(0.1nm,$J_z$=1T,$J_x$=10$^{-2}$T)/*Al*(20nm)/*Be*, curve 2 - *Be*(30nm)/*Al*(57nm)/*Co*(3nm,$M_z$=1T,$M_x$=10$^{-2}$T)/*Al*(20nm)/*Be*, curve 3 – layer *Co*(0.1nm,$M_z$=1T, $M_x$=10$^{-2}$T), curve 4 – layer *Co*(3nm,$M_z$=1T,$M_x$=10$^{-2}$T).

## 4. Conclusion

Thus it has been shown that when spin flip takes place in a resonator resonances of the same order are realized in pairs. If resonances overlap the reflection coefficient and the density coefficient are enhanced as a square function and as a cubic function, respectively, relative to enhancement of nonspin-flip neutron density. This allows carrying out measurements with weakly magnetized and very thing (of an order of an angstroem) layers.

**Acknowledgment**
The work is supported by RBRF grant 08-02-00467-a. Yu.V. Nikitenko thanks V.K. Ignatovich for fruitful discussions and the reviewer for his comments, which has helped us to improve the text considerably.